# Non-local effects in the shear banding of a thixotropic yield stress fluid

M.R. Serial[1], D. Bonn[2], T. Huppertz[3,4], J.A. Dijksman[5], J. van der Gucht[5], J.P.M. van Duynhoven[1,6], C. Terenzi[1]

[1]*Laboratory of Biophysics, Wageningen University & Research, Stippeneng 4, 6708WE, Wageningen, The Netherlands*
[2]*Van der Walls-Zeeman Institute, Institute of Physics, University of Amsterdam, 1098XH Amsterdam, The Netherlands*
[3]*Food Quality and Design, Wageningen University & Research, Stippeneng 4, 6708WE, Wageningen, The Netherlands*
[4]*FrieslandCampina, Stationsplein 4, 3818LE, Amersfoort, The Netherlands*
[5]*Physical Chemistry and Soft Matter, Wageningen University & Research, Stippeneng 4, 6708WE, Wageningen, The Netherlands*
[6]*Unilever Food Innovation Centre Hive, Bronland 14, 6708 WH, Wageningen, The Netherlands*

We observe a novel type of shear banding in the rheology of thixotropic yield-stress fluidsthat is due to the coupling of *both* non-locality and thixotropy. The latter is known to lead to shear banding even in homogeneous stress fields, but the bands observed in the presence of non-local effectsappear different as the shear rate varies continuously over the shear band. Here, we introduce a simple non-local model for the shear banding (NL-SB), and we implement it for the analysis of micron-scale rheo-MRI velocimetry measurements of amilk microgel suspension in a cone-and-plate geometry. The proposed NL-SB model accurately quantifies the cooperativity length and yieldsvalues in the order of the aggregate size in the microgel.

Many yield-stressfluids, such as microgels,are known to exhibit thixotropic behavior, with their viscosity reversibly decreasingwhen the material is subjected to flow. This is due to a competition between the breakdown of the gel structure caused by flow, and its spontaneous build-up at rest. In rheology,one thixotropic signaturecan be found in thehistory- or time-dependent divergence of viscosityat a critical shear rate ($\dot{\gamma}_c$), below which part of the material does not flow. The existence of a critical shear rate then directly implies shear banding, even in the absence of stress variations throughout the gap: for an imposed shear rate $\dot{\gamma} < \dot{\gamma}_c$, only a fraction of the fluid is sheared at $\dot{\gamma}_{loc} = \dot{\gamma}_c$, while the rest remains stagnant ($\dot{\gamma}_{loc} = 0$) [1–4]. Due to the coupling of the microstructure with the flow, the rheological properties of such thixotropic materials are highly sensitive to shear history, which in turn makes quantitatively modelling the flow properties of these systems a yet unresolved challenge. Examples of particulate thixotropic materials,commonly encountered in daily life,include many personal care products [5], food dispersions [6], paints [7], and clay suspensions [8]. In such systems, thixotropy is often the result of weak attractive forces between particles [9]. At rest, or even at low applied shear rates, attractive forces generate the formation of flocs and therefore the development of a particulate network. Inter-particle attractive forces are, however, weak enough to be exceeded by shear forces during flow, thus disrupting the formed network and producing time-dependent behavior.

While a particulatefluid undergoes thixotropic behavior, under strongly-confined flow conditions *cooperative effects*can also become significant, thus introducing an additional level of complexity. Cooperative, or non-local, effects have been observed in emulsions and flowing particle dispersions, when the dimensions of the flow geometrybecome comparable to the characteristic size of the fluid: particles of floc size for particulate gels, or droplets for emulsions [10,11]. The cooperativity effect results from a spatial correlation in the plastic rearrangements induced by local stress fluctuations. A striking consequence of this is that the material's viscosity variesspatially across the confined dimension. Such collective effectshavebeen observed in granular [12,13]or microgel [14–16] suspensions, emulsions [17,18]and cellulose dispersions [19], by means of particle imaging velocimetry (PIV) [11–14,19], optical microscopy [14,15], or rheo-MRI velocimetry [19]. Most experimental observations of non-local effects focusedso far on flow characterization inside narrow microchannels ($50-250$ μm in diameter [11,14,16]),with only one workreporting cooperativity effects in a 1 mm gap cone-and-plate (CP) geometry by rheo-MRI [19].

The emergence and origin of shear banding is one of the most important open questions in the physics of disordered materials, as underlined in authoritative reviews [21,22].Despite the success of the non-local (NL) fluidity model and of related constitutive equations [10,11,20,23] in describing cooperativity effects of a wide range of materials, shear-localization effects are often attributed to,or modelled as being caused by,spatial variations of the stress across the measurement geometry, *e.g.* a microfluidic channel or Couette geometry [20,24]. We show here that in the case of particulate thixotropic fluids, cooperativity and shear localization effects coexist under strongly confined flow, even under homogeneous stress conditions. Even though the rheological properties of thixotropic fluids have been thoroughly studied and modelled for over 50 years [25], confinement effects are commonly overlooked even in rheological tests where gap sizes < 1 mm are used [26]. In our present case, this results in shear bands with a spatially-varying shear rate and, hence, a spatially-varying viscosity. This is in contrast to what happens for 'usual' thixotropic fluids, where two discrete bands are observed, namely stationary or flowing at the critical shear rate, as typical for SB models.

In this work, we perform rheo-MRI velocimetry measurements in a CP geometry, for which the stress is homogeneous to within 1.5%. We demonstrate that the heterogeneous confined flow behavior of acid milk microgel suspensions, known for their thixotropic properties [27,28], can be quantified by accounting for *both* non-locality (NL) and thixotropic shear-banding (SB) within a unified analytical model. By combining the established NL model [11] with the classical SB condition by Møller *et al.* [1], we derive a comprehensive, yet simple, three-parameter fitting expression that successfully describes the spatially-heterogeneous flow of a particulate thixotropic fluid under strong confinement. The proposed NL-SB model enables quantifying cooperativity length scales without any artifact due to concurrent shear-banding effects.

We study the flow behavior of acid milk microgel suspensions (12 wt% milk), using bulk rheology and $^1$H rheo-MRI velocimetry. Acid milk gels are particulate gels; fractal, adhesive hard spheres and percolation models, have been used to model their structure [29–31]. Gels are produced by acidification of milk using 2 wt% of glucono-$\delta$-lactone (GDL) and incubation at 40 °C for 4 hours (pH ~4.2). To ensure reproducible flow behavior, the gels are pre-sheared prior to measurements [32] using a 50 mL syringe pump at a speed of 80 mL/min. The thixotropic properties of the microgel suspension are characterized by measuring hysteresis flow curves using an Anton Paar MCR 301 rheometer equipped with a CP geometry (cone angle 1°, radius 25 mm). The geometry was chosen in order to ensure a homogeneous stress distribution along the cell gap (~ 1.5 %) and to avoid possible stress heterogeneity effects, such as particle-migration, during shearing experiments. Shear rate sweeps for the acid milk microgel suspension, measured by varying the shear rate from 0.01 to 200 $s^{-1}$, and back down to 0.01 $s^{-1}$, are shown in Fig. 1. The applied shear rate is kept constant at each recorded point for 10 seconds. The remarkable difference between the two sweep-rate branches indicates strong thixotropic behavior and a significant stress decay due to structural degradation. Similar results are found for waxy crude oils known for their complex thixotropic behavior [33,34].

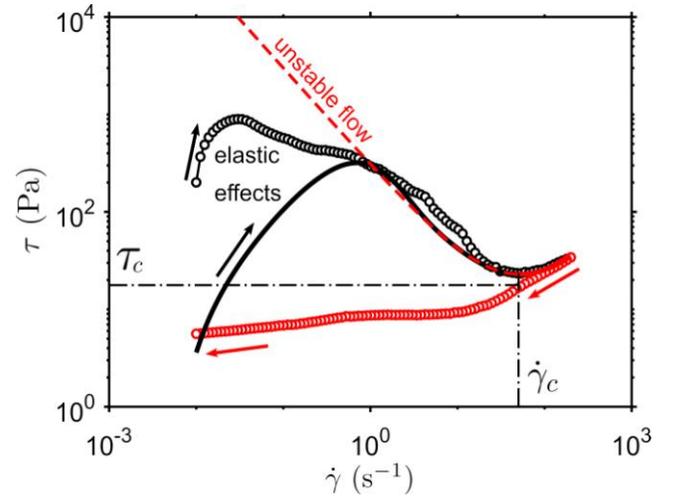

*Fig. 1: Comparison between up (black) and down (red) hysteresis loops obtained from bulk rheology measurements of an acid milk microgel suspension (empty circles), and from the $\lambda$-model (thick lines) with parameters chosen to match the experimental data. The critical shear stress ($\tau_c$), obtained from viscosity bifurcation experiments, and the critical shear rate ($\dot{\gamma}_c$) are indicated as thin dot-dashed lines. The down-sweep curve of the $\lambda$-model is unstable ($d\tau/d\dot{\gamma} < 0$) for $\dot{\gamma} < \dot{\gamma}_c$ (red dashed line).*

In the past years, several models have been proposed to account for thixotropy, with varying degrees of complexity and detail [25]. Here, we consider a minimal model to get some qualitative insight into the system, without accounting for more complicated effects, such as elasticity or kinematic hardening. In the most rudimentary model for thixotropy one assumes that the viscosity $\eta$ increases with dimensionless microstructural parameter $\lambda$ as $\eta = \eta_0 (1 + \lambda^n)$, where $\eta_0$ is the limiting viscosity at high shear rates, and $n$ is a dimensionless parameter specific for the material [4]. For an aging system at rest, or at low shear rates, the structural parameter $\lambda$ increases without bound, while if the structure is disrupted by flow, $\lambda$ decreases to a steady-state value. In this way, the time evolution of the microstructural state of the material can be described as: $\frac{d\lambda}{dt} = \frac{1}{\kappa} - \alpha \lambda \dot{\gamma}$, where $\kappa$ is the characteristic aging time for the microstructure buildup, and $\alpha$ is the rate of disruption. Simple yield-stress behavior is recovered only for $n=1$, while thixotropic effects ($n>1$) are expected to increase with increasing $n$ values [4]. Here we assume $n=2$ as previously done in [2,35] for a similar system. Solving these equations for the dynamic state, and assuming that the shear stress of the material is given by $\tau = \eta \dot{\gamma}$, hysteretic flow curves are obtained that are characteristic of thixotropic systems: because the structure is destroyed by the flow, in an up-and-down shear rate sweep, the downward branch is significantly below the upward branch, as can be observed in Fig.1. The comparison with the experiment in which shear rates is ramped up shows that, although the $\lambda$-model does not account for elastic startup effects [36] visible in the measurements at low shear rates, it qualitatively describes the unstable flow curve with a decreasing stress for increasing shear rate, that is indicative of shear banding. The $\lambda$-model shows an onset of flow instability at $\dot{\gamma}_c \sim 38$ s⁻¹, below which the flow curve has a negative slope. This value defines a critical shear rate above which the viscosity increases to infinity discontinuously [4]. The obtained critical shear rate is also in good agreement with viscosity bifurcation experiments on our milk microgel suspension (Fig.S1 in SupplementalMaterial [37]).

The negative part of the flow curve and associated existence of a critical shear rate automatically imply that, when applying a shear rate smaller than $\dot{\gamma}_c$, shear-banding should be observed [1,3,4]. To look for such behavior in our system, we acquire ¹H velocity profiles with 29 μm spatial resolution by rheo-MRI measurements at 7 T magnetic field using a similar CP geometry (7° cone angle, -1.5 % stress variation and 8 mm radius) as for the rheology measurements [19]. The imposed shear rate is varied between 1 and 50 s⁻¹. To assure reproducible results, the sample is pre-sheared at 30 s⁻¹ for 2 minutes before the shear rate is imposed and a velocity profile is acquired in ~10 minutes.

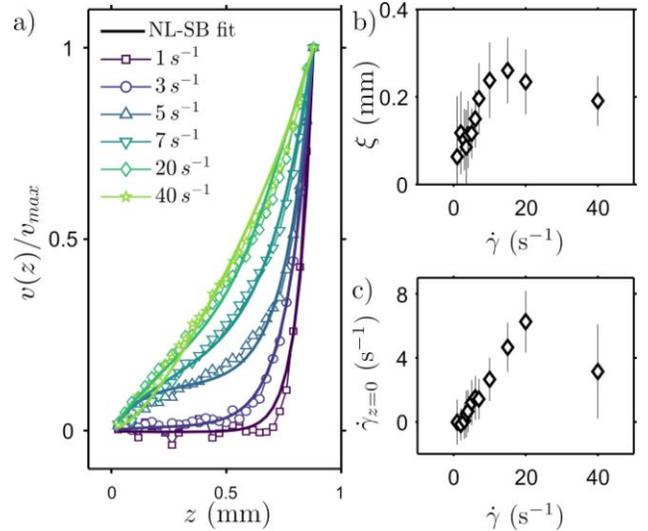

Fig. 2: Fitting results for the ¹H rheo-MRI velocity measurements at increasing applied shear rates, for the acid milk microgel suspension, obtained using the NL-SB model: (a) normalized velocity profiles; (b) $\xi$-values and (c) $\dot{\gamma}_{z=0}$-values. Each rheo-MRI velocity profile was corrected for slippage by extrapolation of the last 5 pixels at the plate, as in [19,20,24].

The flow profiles, plotted in Fig. 2a, show discrete shear bands for the lowest shear rates, $\dot{\gamma} = 1 - 3$ s⁻¹, in agreement with the expected behavior for thixotropic fluids sheared at $\dot{\gamma} < \dot{\gamma}_c$ [1,3,4]. We note that for all rheo-MRI measurements presented here, no time or velocity fluctuations were detected in the static band during the experiment, excluding possible creep effects. Upon increasing the macroscopically imposed shear rate ($\dot{\gamma} > 3$ s⁻¹ in Fig. 2a), a sharp transition from discretely banded to curved velocity profiles is observed. The surprise here is that, in spite of the stress being homogeneous, the flow profiles do not at all have a uniform shear rate. For thixotropic fluids two bands are generally observed below the critical shear rate: one at zero, and one at the critical shear rate. A constant shear rate equal to the imposed shear rate should be observed for $\dot{\gamma} \geq \dot{\gamma}_c$ [1,3,4]. Thus, because of the constant stress, the velocity profiles should always be straight lines. Instead, what we observe is a continuously varying slope of the velocity profiles, implying that the viscosity varies spatially. This is the hallmark of non-local cooperative effects.

To quantitatively account for the coupling between non-locality and shear banding, we consider the non-local fluidity model [11,20], which highlights the effectiveness of the key concept of fluidity, the inverse of the fluid's viscosity $f = \dot{\gamma}/\tau$. During flow, the microstructure of the material causes the fluidity to be influenced by cooperative plastic rearrangements. Thus, by defining a cooperativity length $\xi$ that describes the spatial correlation between plastic deformations, the local fluidity $f(z)$ is set to obey the following diffusion-type equation:

$$\xi^2 \frac{\partial^2 f(z)}{\partial z^2} = f(z) - f_{bulk}(z) \qquad (1)$$

where $f_{bulk}(z)$ corresponds to the fluidity in absence of non-local effects, and $z$ is the position across the cell gap. Under homogeneous flow conditions, Eq. (1) can be solved for a $z$-independent $f_{bulk}$ parameter since $\dot{\gamma}$ and $\tau$ are related through their constitutive equation in the absence of flow cooperativity. However, if the fluid displays shear-banding effects, bulk fluidity is expected to vary within the cell gap. To describe the shear-banding behavior of thixotropic fluids in the absence of non-local effects, we refer to previous pioneering works [1,3,4]. Under homogeneous stress conditions, the velocity of a thixotropic fluid at an imposed shear rate $\dot{\gamma}_{bulk}$ in the absence of non-local effects can be described as:

$$v_{bulk}(z) = \dot{\gamma}_{bulk} H(l-1)z + \\ + \dot{\gamma}_c H(1-l) H(z-lh)(z-(1-l)h) \qquad (2)$$

where $\dot{\gamma}_c$ is the critical shear rate, $l = \dot{\gamma}_{bulk}/\dot{\gamma}_c$ is the fraction of the gap sheared at $\dot{\gamma}_{bulk}$, $h$ is the gap of the geometry, and $H$ is the Heaviside function ($H(x)=0$ for $x<0$ and $H(x)=1$ for $x>0$). From Eq.(2) it follows that: (i) for $\dot{\gamma}_{bulk} < \dot{\gamma}_c$, two shear bands coexist: one static and one sheared at $\dot{\gamma}_{loc} = \dot{\gamma}_c$; (ii) for $\dot{\gamma} \geq \dot{\gamma}_c$, spatially-uniform flow is restored [1]. By considering a $z$-independent stress field $\tau_0$ and solving Eq.(1) for the bulk fluidity taken as $f_{bulk}(z) = \frac{1}{\tau_0}\frac{dv_{bulk}(z)}{dz}$, the following expression for $\dot{\gamma}(z) = \tau_0 f(z)$ is obtained:

$$\dot{\gamma}(z) = A(h,\xi)\cosh\left(\frac{z}{\xi}\right) + B(h,\xi)\sinh\left(\frac{z}{\xi}\right) \qquad (3)$$

$$+ \dot{\gamma}_c (H(l-1)-1) H(z-(1-l)h)$$

$$\times \left(\cosh\left(\frac{z-(1-l)h}{\xi}\right)-1\right) + \dot{\gamma}_{bulk} H(l-1)$$

where $A(h,\xi)$ and $B(h,\xi)$ are $z$-independent constants. We note that Eq.(3) is simply the solution for a homogeneous bulk fluidity plus a second term that depends on the critical value $\dot{\gamma}_c$. By integrating Eq.(3) and applying boundary conditions for the velocity at the cell walls, an analytical equation for the flow profile $v(z)$ can be found. We consider the case of a geometry with homogenous stress field, where one of the cell walls rotates at $\dot{\gamma}_{bulk}$.

This simple model (solid lines in Fig.2a) is in excellent agreement with our rheo-MRI data for the acid milk microgel dispersion. This confirms that the NL-SB model provides a velocity equation $v(z)$ that can be used for quantifying $\xi$ from the experimental rheo-MRI data. The parameter $\dot{\gamma}_c$ is considered a material property and, thus, is here taken as a global fitting parameter for all velocity profiles. Figure 2a shows the fitted rheo-MRI velocities using a three-parameter fitting with the NL-SB model. The fitting parameters $\xi$ and $\dot{\gamma}_{z=0}$ are shown in Figs. 2b and 2c, respectively. Fixed parameters are $v(0) = 0$ (corrected for slippage at the plate), $v(h)$ and the applied shear rate $\dot{\gamma}_{bulk}$. Within the fitting error, the NL-SB model succeeds in describing the spatial dependence of experimental velocities, giving $\xi$-values in the range $50-260$ μm, as expected for similar samples [38] (See Supplemental Material for fitting results for a shear-independent $\xi$ [34]). The observed $\xi$-values are comparable with the aggregate sizes observed in confocal images (Fig.S2 in Supplemental Material [37]). For emulsions, $\xi$-values have been empirically found to depend linearly on the droplet volume fraction above the jamming concentration, and to scale with particle diameter [11]. At the same time, the fitted critical shear rate $\dot{\gamma}_c = (24\pm 3)s^{-1}$ is similar to that obtained from Fig.1; the differences are likely due to reproducibility variations during sample preparation. Fitted $\xi$-values are also comparable with the width of the observed shear band in Fig.2. Similar behavior was found by de Cagny *et al.* for dry granular materials [39].

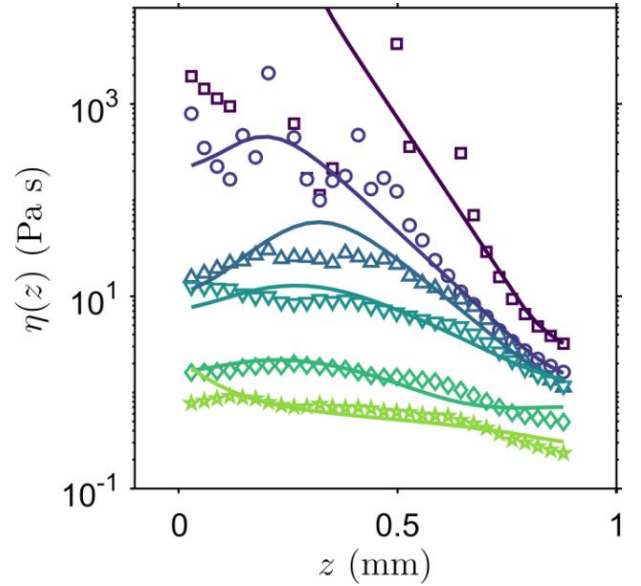

*Fig. 3: Comparison between viscosity profiles calculated from $^1H$ rheo-MRI measurements (symbols) and fitted velocities obtained using the NL-SB model (lines) at increasing shear rate values: 1 s$^{-1}$ (squares), 3 s$^{-1}$ (circles), 5 s$^{-1}$ (triangles), 7 s$^{-1}$ (inverted triangles), 20 s$^{-1}$ (diamonds) and 40 s$^{-1}$ (stars).*

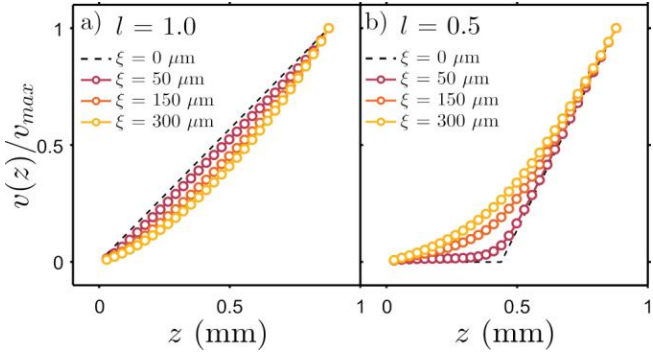

*Fig. 4. Theoretical velocity profiles $v(z)$ obtained with the NL-SB model for different values of the correlation length $\xi$ at (a) $l = 1.0$ and (b) $l = 0.5$, taking into account non-local effects only or coexistent non-local and shear-banding effects, respectively.*

The non-local effects immediately become evident when we construct viscosity profiles (Fig. 3) by combining the globally imposed stress measurements from the rheometer, with the local shear rate from the rheo-MRI (symbols) and fitted NL-SB (lines) velocities. We note that the derivative of the velocity profiles needs to be taken in order to obtain spatial variations of shear rates, and to enhance any possible small deviations between rheo-MRI and NL-SB velocities. The resulting profiles presented in Fig.3 strongly differ from the expected behavior for homogeneous fluids but, in addition, the observed viscosity variations cannot be attributed to non-locality only since this does not account for the observed shear banding. We therefore must conclude that the spatial heterogeneity of the viscosity profiles is caused by the coexistence of *both* non-local and shear-banding effects during flow. Under homogenous flow conditions ($l \geq 1$), non-local effects introduce spatial heterogeneities leading to the observed curved velocity profiles (Fig. 4a). However, for $l < 1$, the cooperativity smoothens the transition between the two discrete bands (Fig. 4b), which becomes more pronounced when increasing $\xi$: for high enough $\xi$-values, the interface between flowing and static bands can no longer be discriminated. We note however that, in such cases, flow profiles for $l > 1$ and $l < 1$ can become very similar (see velocities for $\xi = 300 \mu m$ in Fig.4). This highlights the importance of an NL model that enables disentangling SB from cooperativity effects in SB fluids.

An interesting feature is the dependence of $\xi$ on the applied shear rates seen in Fig. 2b: at low flow rates, $\xi$ increases linearly to $\sim 260$ μm, where it stabilizes for $\dot\gamma \geq 10 \text{ s}^{-1}$. While most experimental results report flow-independent cooperativity lengths [11,14], modelling predicts $\xi$ to increase with the local stress distance to the critical value $\tau_c$, and with the particle size diameter [10,40]. The results presented in Fig.2b for $\dot\gamma < 10 \text{s}^{-1}$ are in agreement with such scaling of $\xi$ when approaching $\dot\gamma_c$: we note that in rheo-MRI experiments shear rate is controlled instead of stress. Yet, flow cooperativity can also be affected by shear-induced density heterogeneities, as previously reported in [13] for a granular material. In the case of attractive gel particles, Rajaram *et al.* [41] recently observed that low shear caused microstructural heterogeneities and pronounced segregation of flocs in a CP geometry by means of 3D confocal imaging. Furthermore, a linear increase of the size of cooperatively reorganizing clusters of flowing soft colloidal microgels was observed by van de Laar *et al.* [16] at increasing microcapillary pressure. The dimensions and density of the formed flocs have been shown to depend on the size of the flow confinement and on a critical shear rate below which attractive forces dominate the system hydro-dynamics [42]. The latter effect could explain the sharp transition observed in our results, where the $\xi$-trend changes from linear to constant behavior at $\dot\gamma = 10 \text{ s}^{-1}$. Flow-concentration coupling effects likely also contribute to the rather complicated spatial variations of the viscosity profiles in Fig.3.

The confinement geometry could also have an impact on the observed flow cooperativity, as previously addressed by Paredes *et al.* [43] in the study of emulsions. The cooperativity, as Goyon, Colin, Ovarlez and collaborators have also shown, is only visible when the system is sufficiently confined. The cooperativity length should be thought of as the size of a cooperatively rearranging region, which in our system may also be seen as a characteristic size of the structure. When, for a given shear rate, the size of the structure becomes comparable to the confinement, cooperative flow effects start to show up. We believe the same to be applicable for our current system.

In conclusion, we have shown that the confined flow of a thixotropic fluid, such as an acid milk microgel suspension, undergoes spatially-heterogeneous flow that cannot be described by modelling non-local flow with the established NL model [20,24] without taking shear banding into account. By introducing a comprehensive NL-SB model, we show that, under confined flow conditions, both effects are strongly intercorrelated, leading to spatial variations of the viscosity. Moreover, by enforcing a spatial dependence in the diffusive equation for the non-local fluidity, we successfully model the flow behavior of the microgel suspension over the full range of applied shear rates, obtaining cooperativity lengths on the same order as the protein aggregate size in the sample. Our results show that cooperativity lengths are not rate-independent, as previously observed in concentrated emulsions [11]. This is not unique to our system; in fact, similar behavior was found by de Kort *et al.* [19] for microfibrillar cellulose dispersions, where the presence of curved velocity profiles under uniform stress conditions was incompletely modelled as caused by non-local effects only. Here, we have shown that shear banding also has to be taken into account.

The observations here are therefore believed to be general for thixotropic systems with a flow-concentration coupling. A more detailed study of the relationship between material properties, such as particle size, attractive or repulsive nature of inter-particle interactions [42], and non-local flow parameters would be interesting future work. We foresee that our proposed NL-SB model can be useful for describing other thixotropic materials flowing in confinement during processing, such as paints, muds, and a range of complex food materials.